# Dipole Anisotropy in the *COBE* [1]-DMR First-Year Sky Maps

A. Kogut[2,6], C. Lineweaver[3], G. F. Smoot[3], C. L. Bennett[4], A. Banday[2],
N. W. Boggess[4], E. S. Cheng[4], G. De Amici[3], D.J. Fixsen[5], G. Hinshaw[2,6],
P. D. Jackson[6], M. Janssen[7], P. Keegstra[6], K. Loewenstein[6], P. Lubin[8],
J.C. Mather[4], L. Tenorio[3], R. Weiss[9], D. T. Wilkinson[10], and E. L. Wright[11]

[1] The National Aeronautics and Space Administration/Goddard Space Flight Center (NASA/GSFC) is responsible for the design, development, and operation of the Cosmic Background Explorer (*COBE*). Scientific guidance is provided by the *COBE* Science Working Group. GSFC is also responsible for the development of analysis software and for the production of the mission data sets.
[2] Universities Space Research Association, Code 685.9, NASA/GSFC, Greenbelt MD 20771
[3] LBL, SSL, & CfPA, Bldg 50-351, University of California, Berkeley CA 94720
[4] NASA Goddard Space Flight Center, Code 685, Greenbelt MD 20771
[5] Applied Research Corporation, Code 685.3, NASA/GSFC, Greenbelt MD 20771
[6] Hughes STX Corporation, Code 685.3, NASA/GSFC, Greenbelt MD 20771
[7] Jet Propulsion Laboratory, Pasadena CA 91109
[8] UCSB Physics Department, Santa Barbara CA 93106
[9] MIT Department of Physics, Room 20F-001, Cambridge MA 02139
[10] Princeton University Department of Physics, Princeton NJ 08540
[11] UCLA Astronomy Department, Los Angeles CA 90024-1562

# ABSTRACT

We present a determination of the cosmic microwave background dipole amplitude and direction from the *COBE* Differential Microwave Radiometers (DMR) first year of data. Data from the six DMR channels are consistent with a Doppler-shifted Planck function of dipole amplitude $\Delta T = 3.365 \pm 0.027$ mK toward direction $(l^{II}, b^{II}) = (264.4° \pm 0.3°, 48.4° \pm 0.5°)$. The implied velocity of the Local Group with respect to the CMB rest frame is $\vec{v}_{LG} = 627 \pm 22$ km s$^{-1}$ toward $(l^{II}, b^{II}) = (276° \pm 3°, 30° \pm 3°)$. DMR has also mapped the dipole anisotropy resulting from the Earth's orbital motion about the Solar system barycenter, yielding a measurement of the monopole CMB temperature $T_0$ at 31.5, 53, and 90 GHz, $T_0 = 2.75 \pm 0.05$ K.

*Subject headings:* cosmic microwave background — cosmology: observations

# 1. Introduction

The cosmic microwave background (CMB) dominates the brightness of the sky at millimeter wavelengths, providing a useful probe of the early universe. The CMB anisotropy on large angular scales consists of a dipole of amplitude $\Delta T/T \sim 10^{-3}$ with quadrupole and higher-order terms two orders of magnitude smaller (Smoot et al. 1992, Bennett et al. 1992b). This anisotropy is usually interpreted in terms of gravitational potential: the quadrupole and higher-order multipoles result from CMB photons leaving regions on the surface of last scattering with different gravitational potentials, while the dipole anisotropy results from the Doppler shift of the solar velocity accumulated under these potentials as they evolved to the present epoch. Measurements of the local peculiar velocity thus provide information on the local distribution of mass in the universe, which can be compared to theoretical models of the growth of structure in the universe (Gorski 1991, Partridge 1988).

Motion with velocity $\vec{\beta} = \vec{v}/c$ through an isotropic radiation field of intensity $I(\nu)$ yields an observed differential intensity

$$\frac{\Delta I}{I}(\nu, \theta) = \frac{I_{obs}(\nu, \theta) - I_{em}(\nu)}{I_{em}(\nu)} \quad (1)$$

where $\nu$ is the frequency in the observer's frame, and $\theta$ is the angle between $\vec{\beta}$ and the direction of observation as measured in the observer's frame. Since $I/\nu^3$ is a Lorentz invariant, the observed Doppler-shifted differential intensity to second order in $\beta$ is

$$\frac{\Delta I}{I} \approx \beta \cos(\theta)(3 - \alpha) + \frac{\beta^2}{2}[2\cos^2(\theta)(6 - 3\alpha + \frac{A}{2}) + (\alpha - 3)] \quad (2)$$

where $\alpha = \frac{\partial ln(I(\nu))}{\partial ln(\nu)}$ and $A = \frac{\nu^2}{I(\nu)} \frac{\partial^2 I(\nu)}{\partial \nu^2}$. For a blackbody at temperature $T_0$, this reduces to the frequency-independent formula for the thermodynamic temperature distribution

$$T(\theta) = \frac{T_0(1-\beta^2)^{1/2}}{1 - \beta \cos(\theta)} \approx T_0[1 + \beta \cos(\theta) + \frac{\beta^2}{2}\cos(2\theta) + O(\beta^3)] \quad (3)$$

(Peebles & Wilkinson 1968). In this paper, we present a precise determination of the CMB dipole amplitude and direction derived from the first year of data taken by the *COBE* Differential Microwave Radiometers (DMR).

## 2. Data Processing and Analysis

The DMR experiment, described by Smoot et al. 1990, provides sensitive maps of the microwave sky at wavelengths 9.5, 5.7, and 3.3 mm (frequencies 31.5, 53, and 90 GHz). There are two nearly independent radiometers (A and B) at each frequency, each of which consists of a Dicke-switched heterodyne receiver fed by a matched antenna pair with 7° FWHM beam. DMR measures the difference in power between two regions of the sky separated by 60°. The combined motions of spacecraft spin, orbit, and orbital precession provide full sky coverage every six months (Boggess et al. 1992).

On-board noise sources transfer pre-flight laboratory calibration to orbit (Bennett et al. 1992a); the calibrated time-ordered temperature differences are in units of antenna temperature. Differences in antenna temperature at observing frequency $\nu$ are related to differences in thermodynamic temperature by $\Delta T_A = \Delta T \frac{x^2 e^x}{[e^x-1]^2}$, where $x = h\nu/kT$, $h$ is Planck's constant, and $k$ is Boltzmann's constant. We adjust the calibration in the 31B channel by $+4.9\%$ and the 90B channel by $-1.4\%$ based on observations of the Moon, to force identical lunar brightness in the A and B channels at each frequency. Data in this paper are from the period 22 Dec 1989 through 21 Dec 1990, excluding data taken when the Earth is 1° below the Sun/Earth shield or higher, when the Moon is within 25° of an antenna beam center, when any datum deviates from the daily mean by more than $5\sigma$, or when the spacecraft telemetry or attitude solution is of poor quality. We correct the remaining data for emission from the Moon and Jupiter, the instrument susceptibility to the Earth's magnetic field, and the Doppler effect of the spacecraft velocity about the Earth and the Earth motion about the Solar system barycenter. The largest correction is for the magnetic response, which if uncorrected would create a dipole pattern aligned with the Earth's magnetic field. Residual uncertainties after correction are small (see Table 11 of Kogut et al. 1992).

Two different methods quantify the large-scale anisotropy of the sky antenna temperature. A least-squares minimization fits the time-ordered data directly to a spherical harmonic expansion on the sky, rejecting data with Galactic latitude $|b| < 10°$. A sparse-matrix implementation of a least-squares minimization (Janssen & Gulkis 1991) fits the time-ordered data to sky maps of 6144 pixels using the quadrilateralized spherical cube representation (O'Neill & Laubscher 1976); the sky maps are then fitted to a spherical harmonic expansion using a variety of Galactic cuts. The two methods agree within 3 $\mu$K rms; results in this paper use the mean of the two methods with the difference included as part of the upper limits on combined systematic uncertainty.

The temperature distribution at each frequency is dominated by a dipole anisotropy. Figure 1 shows the $|b| > 10°$ data in the 53 GHz (A+B)/2 summed maps, weighted by the number of observations in each pixel and binned by angle relative to the fitted dipole axis. Deviations from a pure dipole are less than 2% of the dipole amplitude, and are consistent with estimates of higher-order CMB anisotropy and Galactic emission (Smoot et al. 1992, Bennett et al. 1992b). The residuals are strongly correlated between the DMR channels and show the distribution expected for Galactic emission near the plane of the Galaxy. For $|b| > 10°$ the rms of the residuals has a frequency dependence $\nu^p$ with $p = -0.3 \pm 0.7$, as expected from an admixture of Galactic emission and CMB anisotropy (all uncertainties in this paper are 68% confidence level). Determination of the dipole amplitude and direction at the 1% level must account for Galactic emission and the aliasing of higher-order power.

Table 1 lists the observed dipole parameters for $|b| > 10°$ for each of the six DMR channels. The first row under each channel ("Mean") shows the mean value and the last row ("Total Error") the associated 68% CL uncertainty for the Cartesian components in Galactic coordinates, the amplitude, and the direction. All temperatures in Table 1 have been increased by 0.5% to account for the smoothing of the dipole anisotropy by the 7° DMR beam width. The second, third, and fourth rows show the contribution to the total uncertainty caused by the instrument noise, the absolute calibration uncertainty, and the combined upper limits to possible systematic artifacts. The $\sim 1\%$ uncertainty in the absolute calibration dominates the uncertainty in the amplitudes but does not affect the dipole direction. The systematic uncertainty includes the effect of a 10° cut in Galactic latitude, which couples the components of a spherical harmonic decomposition. This effect is small (less than 5 $\mu$K); the systematic uncertainty is dominated by uncertainties in the instrument response to the Earth's magnetic field (Kogut et al. 1992). The antenna pointing was measured before launch and confirmed in flight using observations of the Moon. Uncertainties in the pointing are less than 0.05° at 68% CL (Bennett et al. 1992a, Kogut et al. 1992).

The observed dipole at each frequency is the vector sum of the CMB dipole and a small Galactic term. Figure 2 shows the fitted dipole parameters for the 53 GHz (A+B)/2 sum map as a function of the cut in Galactic latitude. Galactic emission on large angular scales is predominantly quadrupolar, changing the fitted dipole amplitude less than 20 $\mu$K even when no cut is made. The primary effect of Galactic emission is on the dipole direction, shifting the fitted direction $\sim 1°$ toward the Galactic center as data closer to the plane are included in the fit.

We estimate the Galactic contribution to the observed dipole at each frequency using the Galaxy models described in Bennett et al. 1992b. A least-squares algorithm fits the six DMR maps (corrected for synchrotron and dust emission) to free-free and CMB emission of unknown amplitude and fixed frequency dependence using a spherical harmonic basis to order $\ell=2$. Table 2 lists the Galactic dipole components in antenna temperature at 53 GHz. The amplitudes at 31.5 GHz can be derived by multiplying these values by [5.02, 3.05, 0.42] for synchrotron, free-free, and dust emission, respectively; the corresponding ratios for 90 GHz are [0.19, 0.32, 2.34]. The modelled Galactic emission constitutes less than 2% of the observed dipole amplitude at $|b| > 10°$ in any of the DMR maps.

The dipoles in Table 1, corrected for Galactic emission using Table 2, are consistent with a cosmic origin. The amplitudes, expressed in thermodynamic temperature, have spectral index $\nu^p$ with $p = 0.004 \pm 0.017$, where $p = 0$ for a Doppler-shifted Planck spectrum. The fitted CMB dipole amplitude is $\Delta T = 3.365 \pm 0.027$ mK toward $(l^{II}, b^{II}) = (264.4° \pm 0.3°, 48.4° \pm 0.5°)$, assuming a CMB temperature $T_0 = 2.73$ K in the conversion from measured antenna temperature to thermodynamic temperature. The uncertainties in the CMB dipole direction are dominated by the uncertainty in the correction for diffuse Galactic free-free emission, which is limited by the instrument noise in the longest wavelength channels.

Figure 3 shows the dipole amplitude and direction derived from the DMR data and other precise full-sky surveys (Fixsen et al. 1993 [FIRAS], Strukov et al. 1988 [RELIKT], Lubin et al. 1985 [Berkeley], Fixsen et al. 1983 [Princeton]). The DMR dipole agrees with other measurements, in particular that of *COBE*-FIRAS. The FIRAS instrument on *COBE* measures the spectrum of the sky from 60 GHz to 600 GHz. The spectrum at each pixel is fitted to a monopole, dipole, and Galactic dust model, with resulting dipole amplitude $\Delta T = 3.347 \pm 0.008$ mK toward $(l^{II}, b^{II}) = (265.6° \pm 0.8°, 48.3° \pm 0.5°)$ (Fixsen et al. 1993). The FIRAS amplitude reported here has been increased by 0.1% to account for the smearing of the FIRAS beam. Various determinations of the dipole direction are in agreement across two decades of frequency and support the interpretation of the dipole as cosmic in origin. The 0.8° difference between the DMR and FIRAS dipole directions is not statistically significant compared to their uncertainties. The DMR dipole direction is displaced $\sim 1°$ in Galactic longitude away from the Galactic center compared to the other measurements in Figure 3. This is comparable to the change in fitted direction as the Galactic cut angle is varied (Fig 2), and may well represent the different Galactic removal strategies adopted by various groups. The DMR frequencies span the minimum in the Galactic emission spectrum; the DMR dipole parameters are insensitive to the exact choice of cut angle and Galactic model. The uncertainties associated with chosen Galactic emission model and cut angle are explicitly included in the DMR values, and dominate the directional uncertainty.

DMR also provides a measurement of the monopole CMB temperature. Although DMR measures differences in temperature and is not directly sensitive to a monopole, equation (3) can be used to infer the temperature $T_0$ from the dipole spectrum or (provided $\beta$ is known) amplitude. We assume a cosmic origin and fit the Galaxy-corrected dipoles (in antenna temperature) to the derivative of a Planck spectrum to derive $T_0$ and $\beta$, independent of any other data. The derived values are $T_0 = 2.76 \pm 0.18$ K and $\beta = (1.22 \pm 0.09) \times 10^{-3}$. DMR observations of the Doppler shift from the Earth's orbital motion provide a second, independent measurement of $T_0$. In this case, $\beta = 1.01 \times 10^{-4}$ is precisely known. Figure 4 shows the DMR 53 GHz (A+B)/2 data, without the correction for the Earth Doppler shift, mapped in a coordinate system fixed with respect to the Earth orbital velocity. The data follow the expected dipole pattern. The amplitude is directly proportional to $T_0$; the average of the six DMR channels yields $T_0 = 2.75 \pm 0.05$ K ($\chi^2 = 4.4$ for 5 DOF). Both determinations of $T_0$ agree with the much more precise FIRAS measurement $T_0 = 2.726 \pm 0.005$ K (Mather et al. 1993).

## 3.  Discussion

The CMB dipole is usually interpreted as the leading term of the kinetic anisotropies resulting from the Local Group's motion with respect to the rest frame of the CMB. In this scenario, matter, averaged over the largest scales, is at rest with respect to the CMB; bulk motion of horizon-sized volumes is negligible. The Local Group's peculiar velocity has been gravitationally induced since decoupling by local ($< 20,000$ km s$^{-1}$) inhomogeneities in the density field. This interpretation is simple, testable, and consistent with observations.

The frequency independence of the DMR and FIRAS dipole results, as well as the 120:1 ratio of dipole to quadrupole amplitudes, supports the interpretation of a Doppler-shifted Planckian origin. For $T_0 = 2.73$ K, the derived value for the Solar speed is $\beta = (1.23 \pm 0.01) \times 10^{-3}$, or $v = 370 \pm 3$ km s$^{-1}$. We use the velocities listed in Table 3 to derive the Local Group velocity with respect to the CMB, $\vec{v}_{LG} = 627 \pm 22$ km s$^{-1}$ toward $(l^{II}, b^{II}) = (276° \pm 3°, 30° \pm 3°)$.

The Local Group velocity derived from the CMB dipole may be compared to independent determinations of large-scale motions. Spectroscopic observations suggest that the Galaxy, along with other local galaxies, is participating in a bulk flow with a velocity roughly consistent with that inferred from the CMB dipole (Bertschinger et al. 1990, Strauss et al. 1992). The alignment between the apex of the CMB dipole and the calculated dipole range from $\sim 7°$ to $\sim 30°$ (Scaramella et al. 1991, Lynden-Bell et al. 1988, 1989) depending on the method and sample, with error bars of the same magnitude. This fair alignment supports not only the notion that the dipole is kinetic but also that it is gravitationally induced.

Independent but weak confirmation of the standard kinetic interpretation of the CMB dipole comes from a marginal X-ray anisotropy detection by *HEAO-1* (Boldt 1987). The direction of the asymmetry coincides with the CMB dipole and the magnitude is at the expected level (cf equation (2) with $\alpha \approx -1$). The direction of the best fit dipole in the 2-10 keV band is $(l^{II}, b^{II}) = (280°, 30°)$, within 2 $\sigma$ of the CMB dipole (Fabian et al. 1992). The inferred speed of $475 \pm 165$ km s$^{-1}$ is consistent with our reported value $370 \pm 3$ km s$^{-1}$.

Peculiar velocities $v$ with respect to the smooth Hubble flow will result from the effect of primordial gravitational potential gradients integrated over time. DMR measures the power spectrum of large-scale temperature fluctuations, which are proportional to the potential fluctuations, $\frac{\Delta T}{T} = \frac{\Delta \Phi}{3c^2}$ (Sachs & Wolfe 1967). For a random field of pertubations with potential power spectrum $\Phi_k^2$, the resulting mean square peculiar velocity is $<v^2> = 4\pi f^2 \int dk\ k\ \Phi_k^2\ W^2(k)$ or expressed in terms of the density power spectrum $P(k)$, $<v^2> = 4\pi f^2 \int dk\ P(k)\ W^2(k)$ where $W(k)$ is the observation window function and $f$ is the growth factor of the growing mode, $\frac{d\ lnD}{dt} = H_0 \Omega^{0.6}$. For density power spectrum $P(k) \propto k^n$ (potential $\Phi_k^2 \propto k^{n-1}$), the contribution to the rms peculiar velocity $<v^2>^{\frac{1}{2}}$ arising from primordial potential fluctuations of length scale larger than $\lambda$ is

$$<v^2>^{\frac{1}{2}} \approx\ 400\ \text{km s}^{-1}\ \frac{Q_{rms-ps}}{16\ \mu\text{K}}\ [\frac{50h^{-1}\ \text{Mpc}}{\lambda}]^{\frac{n+1}{2}} \quad (4)$$

where $h$ is the Hubble constant $H_0$ in units 100 km s$^{-1}$ Mpc$^{-1}$ (cf. Abbott & Wise 1984). Although the primordial perturbation spectrum likely extended down to scales much smaller than 1 Mpc, the dissipative effects of dark matter and virialization from non-linear processes dominate at scales of 10 Mpc and might intervene to scales of 30 or 40 Mpc. These effects will on average increase the velocity ($v^2$ is positive definite), resulting in a firm lower limit. Since $v^2$ is broadly distributed, the observed value of $627 \pm 22$ km s$^{-1}$ is not atypically large. A specific prediction of these observations is that the rms flow velocity will decrease with sample size $\lambda$ according to equation (4).

The analysis above assumes that the entire CMB dipole is kinetic in origin. In general, the cosmic dipole could be the superposition of kinetic and intrinsic terms. Curvature fluctuations on super-horizon scales will not create an intrinsic dipole anisotropy as both the observer and the CMB photons are in free-fall. If the large-scale CMB anisotropy results from other causes (e.g. isocurvature perturbations), the intrinsic term must be considered. For a density perturbation power spectrum $P(k) \sim k^n$, the power in multipole moment $\ell$ of the CMB temperature distribution is

$$< \Delta T_\ell^2 >_n = Q_{rms\_ps}^2 \frac{2\ell+1}{5} \frac{\Gamma(\ell+\frac{n-1}{2})\Gamma(\frac{9-n}{2})}{\Gamma(\ell+\frac{5-n}{2})\Gamma(\frac{3+n}{2})}. \quad (5)$$

(Bond & Efstathiou 1987). The large-scale CMB anisotropy has $n = 1.1 \pm 0.5$ and $Q_{rms\_ps} = 16 \pm 4$ $\mu$K (Smoot et al. 1992), leading to an "intrinsic" dipole power $< \Delta T_1^2 > \approx 431 \pm 413$ $\mu$K$^2$ (dipole amplitude $36^{+14}_{-29}$ $\mu$K). In such cosmological models, $\sim 1\%$ of the CMB dipole is intrinsic. Analysis of the kinetic component must then include an additional uncertainty $\sim 3$ km s$^{-1}$ in amplitude and $\sim 0.5°$ in direction to account for the unmeasured intrinsic component.

Alternatives to the standard model exist. One hypothesis is that the CMB is not the rest frame of the matter. In this scenario, the matter and photons shared a common bulk velocity at decoupling. Since that time, $v_{matter}$ has scaled as $(1 + z)$ while the photon "bulk" velocity has remained constant. This model predicts that bulk motions extend over the entire horizon (Matzner 1980, Gunn 1988, but see also Burstein et al. 1990). A second alternative proposes that the dipole is due primarily to an entropy gradient (super-horizon isocurvature fluctuation) intrinsic to the surface of last scattering. Radial symmetry (Paczynski & Piran 1990) and single mode fluctuations (Turner 1991) are invoked to explain the small quadrupole to dipole ratio. An alignment of the X-ray dipole (sources at $z \sim 0.1$) with the CMB dipole is evidence against these models.

The standard kinetic interpretation is falsifiable. Critical tests include: 1) In all diffuse backgrounds (X-ray, FIR, gamma-ray and ultra-high-energy cosmic ray) one should detect kinetic dipoles aligned with the CMB dipole with amplitudes depending on the slope $\alpha(\nu)$ of the spectral intensity $I(\nu)$. 2) As bulk motion and cumulative acceleration studies obtain better and deeper galaxy samples, the convergence of the calculated Local Group acceleration vector with the CMB dipole direction should improve from the present $\sim 20°$ discrepancy. 3) As the region of bulk motion observations gets larger, the bulk velocities should approach zero.

## 4. Conclusions

DMR observes dipoles with high signal to noise ratio in each of the six channels. The dipoles are consistent with a Doppler-shifted Planck origin, with amplitude $\Delta T = 3.365 \pm 0.027$ mK toward direction $(l^{II}, b^{II}) = (264.4° \pm 0.3°, 48.4° \pm 0.5°)$. The amplitude uncertainty is dominated by the absolute calibration uncertainty, while the direction uncertainties result from uncertainty in the diffuse Galactic emission. The derived velocity of the Local Group with respect to the CMB is $\vec{v}_{LG} = 627 \pm 22$ km s$^{-1}$ toward $(l^{II}, b^{II}) = (276° \pm 3°, 30° \pm 3°)$. The agreement in dipole amplitude and spectrum between various observations over more than two decades of frequency, the large ratio of dipole to quadrupole amplitudes, and the agreement between the observed Local Group velocity and estimates based on large-scale structure all support a kinematic, gravitationally-induced origin to the CMB dipole.

We gratefully acknowledge the efforts of those contributing to the *COBE* DMR, including J. Aymon, E. Kaita, V. Kumar, R. Kummerer, and J. Santana. *COBE* is supported by the Office of Space Sciences and Applications of NASA Headquarters.

# References

Abbott, L.F. & Wise, M.B. 1984, ApJ, 282, L47
Bennett, C.L., et al. 1992a, ApJ, 391, 466
Bennett, C.L., et al. 1992b, ApJ, 396, L7
Bertschinger, E., Dekel, A., Faber, S.M., Dressler, A., & Burstein, D. 1990, ApJ, 364, 370
Boggess, N.W., et al. 1992, ApJ, 397, 420
Boldt, E. 1987, Physics Reports, 146, 215
Bond, J.R., & Efstathiou, G. 1987, MNRAS, 226, 655
Burstein, D., Faber, S.M., & Dressler, A. 1990, ApJ, 354, 18
Fabian, A.C., Barcons, X. 1992, Ann Rev Ast Astrophys, 30, 429
Fich, M., Blitz, L., & Stark, A. 1989, ApJ, 342, 272
Fixsen, D.J, Cheng, E.S., & Wilkinson. D.T. 1983, PRL, 50, 620
Fixsen, D.J. et al. 1993, ApJ, (submitted)
Haslam, C.G.T., Klein, U., Salter, C.J., Stoffel, H., Wilson, W.E., Cleary, M.N., Cooke, D.J., & Thomasson, P. 1981, A&A, 100, 209
Gorski, K. 1991, ApJ, 370, L5
Gunn, J.E. 1988, The Extra-Galactic Distance Scale, ASP Conf Ser 4, ed. S.van den Bergh & C.J. Pritchet, 344
Janssen, M.A. & Gulkis, S. 1991, in The Infrared and Submillimetre Sky After *COBE*, ed. M. Signore & C. Dupraz (Dordrecht:Kluwer), 391
Kerr, F.J. & Lynden-Bell, D. 1986, MNRAS, 221, 1023
Kogut, A., et al. 1992, ApJ, 401, 1
Lubin, P.M., Villela, T., Epstein, G.L., & Smoot, G.F. 1985, ApJ, 281, L1
Lynden-Bell, D., et al. 1988, ApJ, 326, 19
Lynden-Bell, D., Lahav, O., & Burstein, D. 1989, MNRAS, 241, 325
Mather, J.C., et al. 1993, ApJ, (submitted)
Matzner, R.A. 1980, ApJ, 241, 851
O'Neill, E.M., & Laubscher, R.E., 1976, NEPRF Technical Report 3-76
Paczynski,B. & Piran, T. 1990, ApJ, 364, 341
Partridge, R.B. 1988, Rep. Prog. Phys, 51, 647
Peebles, P.J.E., & Wilkinson, D.T. 1968, Phys Rev, 174, 2168
Reich, P., & Reich, W. 1988, A&AS, 74, 7
Sachs, R.K. & Wolfe, A.M. 1967, ApJ, 147, 73
Scaramella, R., Vettolani, G., & Zamorani, G. 1991, ApJ, 376, L1
Smoot, G.F., et al. 1990, ApJ, 360, 685
Smoot, G.F., et al. 1992, ApJ, 396, L1
Strauss, M. et al. 1992, ApJ, 397, 395
Strukov, I.A. & Skulachev, D.P. 1988, Astrophysics & Space Physics Review, 6, 147
Turner,M. 1991, Fermilab pub-91/43-A
Wright, E.L. et al. 1991, ApJ, 381, 200
Yahil,A., Tammann,G.A. & Sandage,A. 1977, ApJ, 217, 903

Table 1: Observed Dipole Parameters[a] and Uncertainties, $|b| > 10°$

| Channel | Term | $T_X$ ($\mu$K) | $T_Y$ ($\mu$K) | $T_Z$ ($\mu$K) | Amplitude ($\mu$K) | $l^{II}$ (deg) | $b^{II}$ (deg) |
|---|---|---|---|---|---|---|---|
| 31A | Mean | -200 | -2216 | 2406 | 3277 | 264.82 | 47.25 |
| | Noise | 21 | 31 | 23 | 27 | 0.56 | 0.49 |
| | Gain | 5 | 55 | 60 | 57 | 0.00 | 0.00 |
| | Systematics | 16 | 22 | 14 | 18 | 0.43 | 0.34 |
| | Total Error | 27 | 67 | 66 | 66 | 0.71 | 0.60 |
| 31B | Mean | -190 | -2180 | 2396 | 3245 | 265.00 | 47.60 |
| | Noise | 24 | 35 | 26 | 31 | 0.65 | 0.56 |
| | Gain | 4 | 50 | 55 | 52 | 0.00 | 0.00 |
| | Systematics | 21 | 29 | 27 | 28 | 0.56 | 0.50 |
| | Total Error | 32 | 68 | 67 | 67 | 0.86 | 0.75 |
| 53A | Mean | -198 | -2082 | 2314 | 3120 | 264.56 | 47.89 |
| | Noise | 7 | 10 | 8 | 9 | 0.21 | 0.18 |
| | Gain | 1 | 14 | 16 | 15 | 0.00 | 0.00 |
| | Systematics | 9 | 17 | 10 | 13 | 0.25 | 0.27 |
| | Total Error | 11 | 24 | 21 | 22 | 0.33 | 0.32 |
| 53B | Mean | -199 | -2067 | 2353 | 3139 | 264.48 | 48.56 |
| | Noise | 8 | 12 | 9 | 10 | 0.23 | 0.20 |
| | Gain | 1 | 14 | 16 | 15 | 0.00 | 0.00 |
| | Systematics | 7 | 11 | 10 | 10 | 0.22 | 0.20 |
| | Total Error | 11 | 22 | 21 | 21 | 0.31 | 0.29 |
| 90A | Mean | -180 | -1820 | 2058 | 2753 | 264.33 | 48.37 |
| | Noise | 13 | 19 | 15 | 17 | 0.42 | 0.37 |
| | Gain | 3 | 36 | 41 | 39 | 0.00 | 0.00 |
| | Systematics | 8 | 17 | 11 | 14 | 0.27 | 0.32 |
| | Total Error | 16 | 44 | 45 | 45 | 0.50 | 0.49 |
| 90B | Mean | -174 | -1830 | 2029 | 2738 | 264.56 | 47.82 |
| | Noise | 9 | 13 | 10 | 12 | 0.29 | 0.26 |
| | Gain | 2 | 23 | 26 | 25 | 0.00 | 0.00 |
| | Systematics | 6 | 13 | 11 | 12 | 0.20 | 0.26 |
| | Total Error | 11 | 30 | 30 | 30 | 0.35 | 0.37 |

[a] All temperatures are in units of antenna temperature. No corrections have been made for Galactic emission. We express the dipole as $T(l,b) = T_X \cos(l)\cos(b) + T_Y \sin(l)\cos(b) + T_Z \sin(b)$.

Table 2: Galactic Dipole[a] at 53 GHz, $|b| > 10°$

| Emission | $T_X$ ($\mu$K) | $T_Y$ ($\mu$K) | $T_Z$ ($\mu$K) | Amplitude ($\mu$K) | $l^{II}$ (deg) | $b^{II}$ (deg) |
|---|---|---|---|---|---|---|
| Synchrotron | $3.8 \pm 1.2$ | $1.2 \pm 0.4$ | $-1.5 \pm 0.5$ | $4.3 \pm 1.1$ | $18 \pm 8$ | $-21 \pm 8$ |
| Free-Free | $-1.3 \pm 8.7$ | $-8.1 \pm 21.0$ | $-11.6 \pm 20.8$ | $14.2 \pm 20.8$ | $261 \pm 64$ | $-55 \pm 84$ |
| Dust | $0.3 \pm 0.1$ | $0.3 \pm 0.1$ | $-0.2 \pm 0.1$ | $0.5 \pm 0.1$ | $45 \pm 13$ | $-25 \pm 9$ |
| Combined | $2.8 \pm 8.8$ | $-6.6 \pm 21.0$ | $-13.3 \pm 20.8$ | $15.1 \pm 20.5$ | $293 \pm 92$ | $-62 \pm 75$ |

[a] All temperatures are in units of antenna temperature.

Table 3: Relative Velocities

| Description | Velocity (km s$^{-1}$) | $l^{II}$ (deg) | $b^{II}$ (deg) | Reference |
|---|---|---|---|---|
| Sun – CMB | $369.5 \pm 3.0$ | $264.4 \pm 0.3$ | $48.4 \pm 0.5$ | Kogut et al. 1993 (this work) |
| Sun–LSR | $20.0 \pm 1.4$ | $57 \pm 4$ | $23 \pm 4$ | Kerr & Lynden-Bell 1986 |
| LSR–GC | $222.0 \pm 5.0$ | $91.1 \pm 0.4$ | 0 | Fich, Blitz & Stark 1989 |
| GC–CMB | $552.2 \pm 5.5$ | $266.5 \pm 0.3$ | $29.1 \pm 0.4$ | |
| Sun–LG | $308 \pm 23$ | $105 \pm 5$ | $-7 \pm 4$ | Yahil et al. 1977 |
| LG–CMB | $627 \pm 22$ | $276 \pm 3$ | $30 \pm 3$ | |

(LSR: Local Standard of Rest    GC: Galactic Center    LG: Local Group)

# Figure Captions

Figure 1. 53 GHz (A+B)/2 sum map binned by angle relative to the fitted dipole axis. The data have been corrected for the 0.5% smoothing of the DMR beam. (Top) Binned data $|b| > 10°$, with fitted cosine overlay. (Bottom) Residuals after removing fitted dipole. The solid line shows the residuals for $|b| > 10°$. The dashed line shows residuals with no Galactic cut. Deviations from a pure dipole are dominated by Galactic emission and are less than 2% of the dipole amplitude for $|b| > 10°$.

Figure 2. 53 GHz (A+B)/2 fitted dipole parameters as a function of cut in Galactic latitude $|b|$. The amplitude is in units of antenna temperature and has been corrected for the 0.5% smoothing of the DMR beam. All uncertainties are statistical only. The Galactic emission is predominantly quadrupolar. A precise determination of the dipole amplitude and direction must consider Galactic emission.

Figure 3. Various measurements of the dipole amplitude and direction. (Top) Dipole amplitude in thermodynamic temperature. Antenna temperatures have been converted to thermodynamic temperature assuming a 2.73 K blackbody. The DMR data have been corrected for Galactic emission. (Bottom) Dipole direction in Galactic coordinates. The DMR point is the mean CMB dipole derived by fitting the six channels to CMB and Galactic emission.

Figure 4. 53 GHz (A+B)/2 summed data with $|b| > 10°$, mapped in a coordinate system fixed with respect to the Earth velocity about the Solar system barycenter. The data have been binned by angle relative to the Earth velocity vector and corrected for the 0.5% smoothing of the DMR beam. The amplitude of the dipole is proportional to $T_0$ and $\beta$ (Eq. 3).

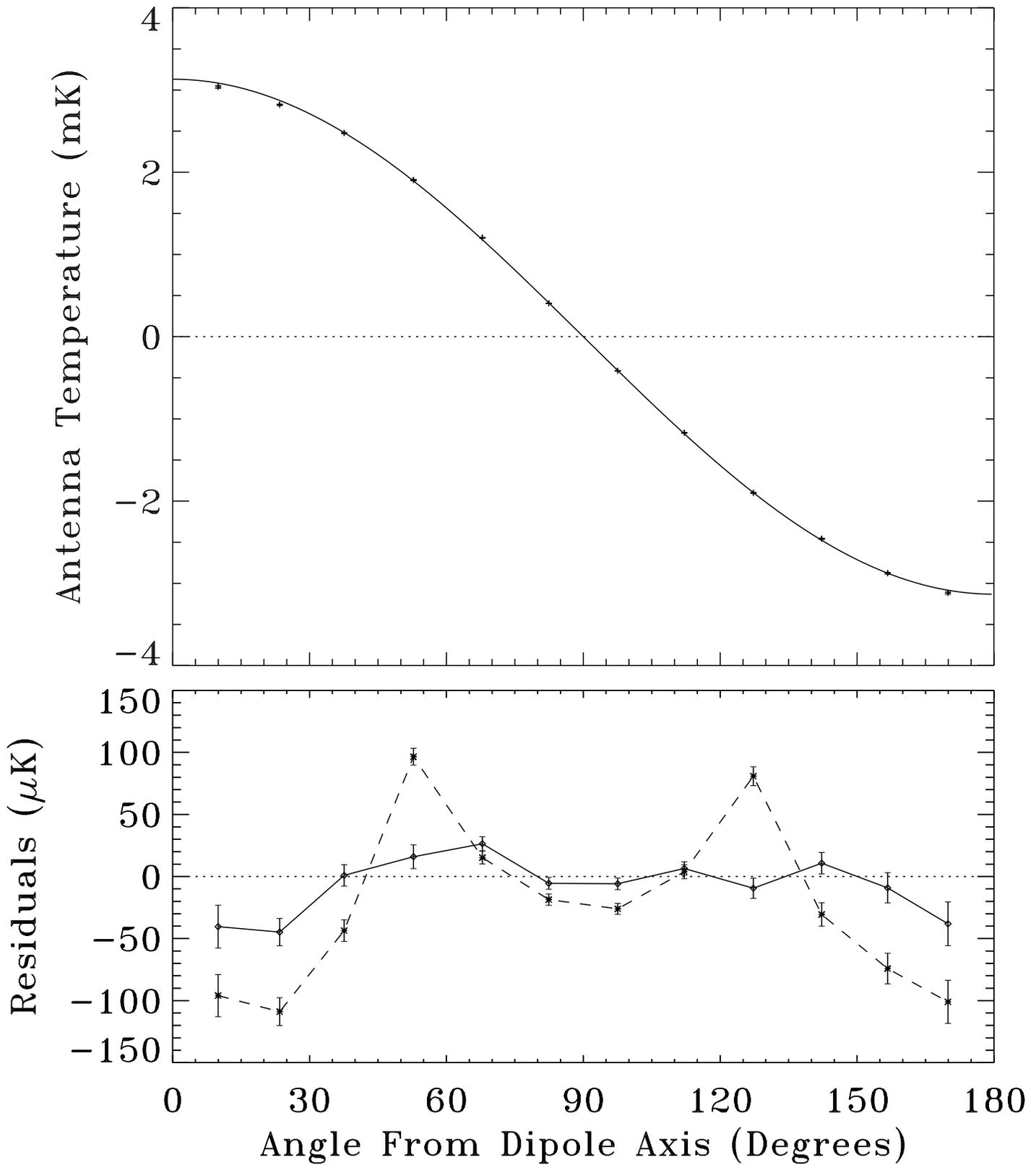

Figure 1

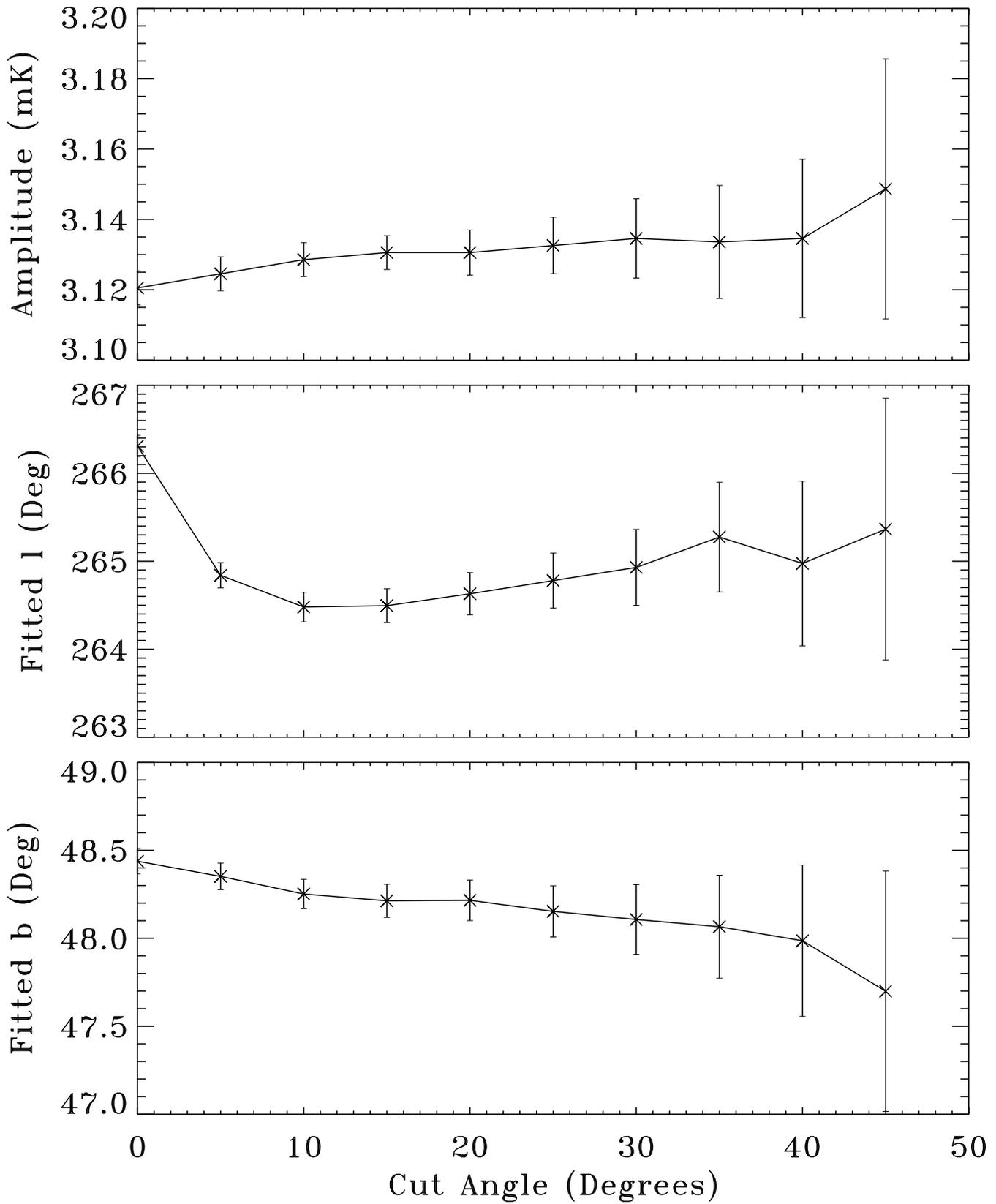

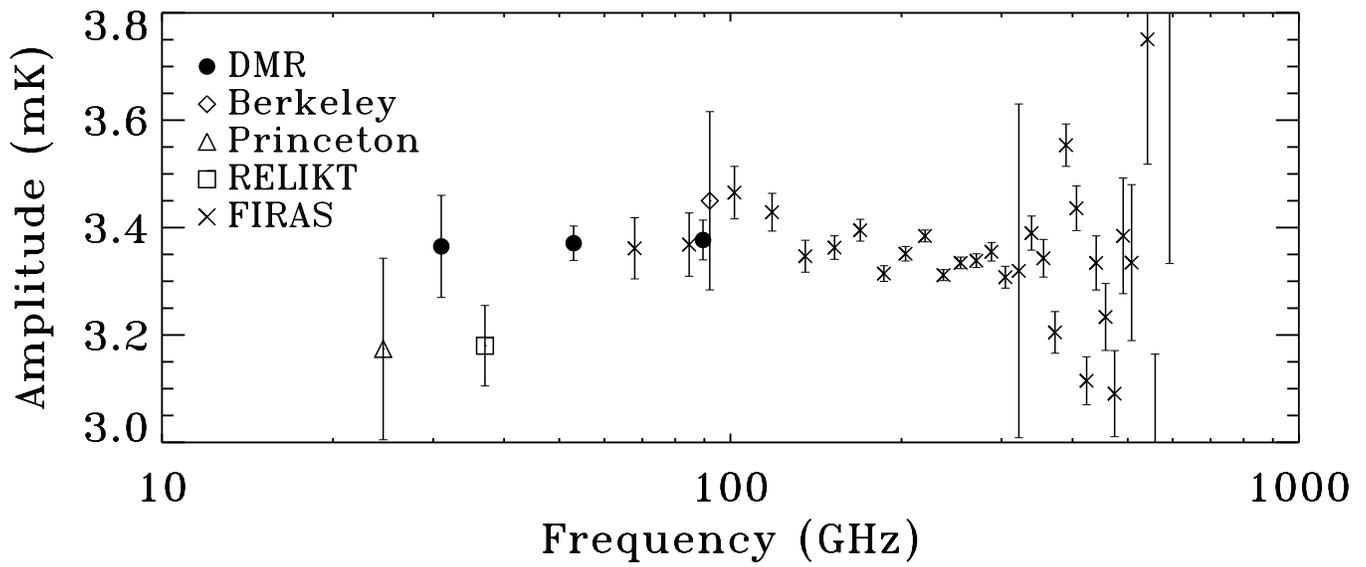
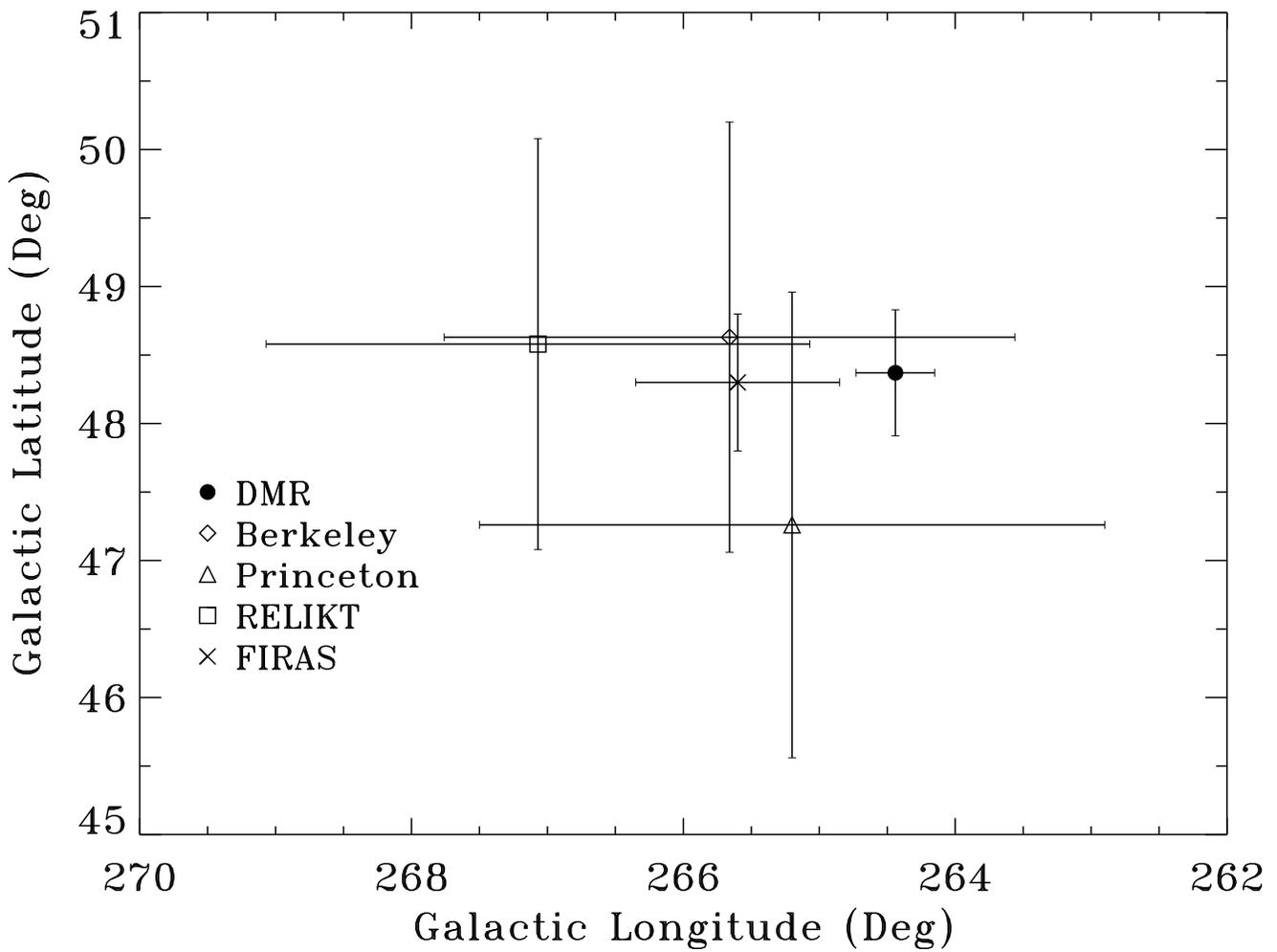

Figure 3

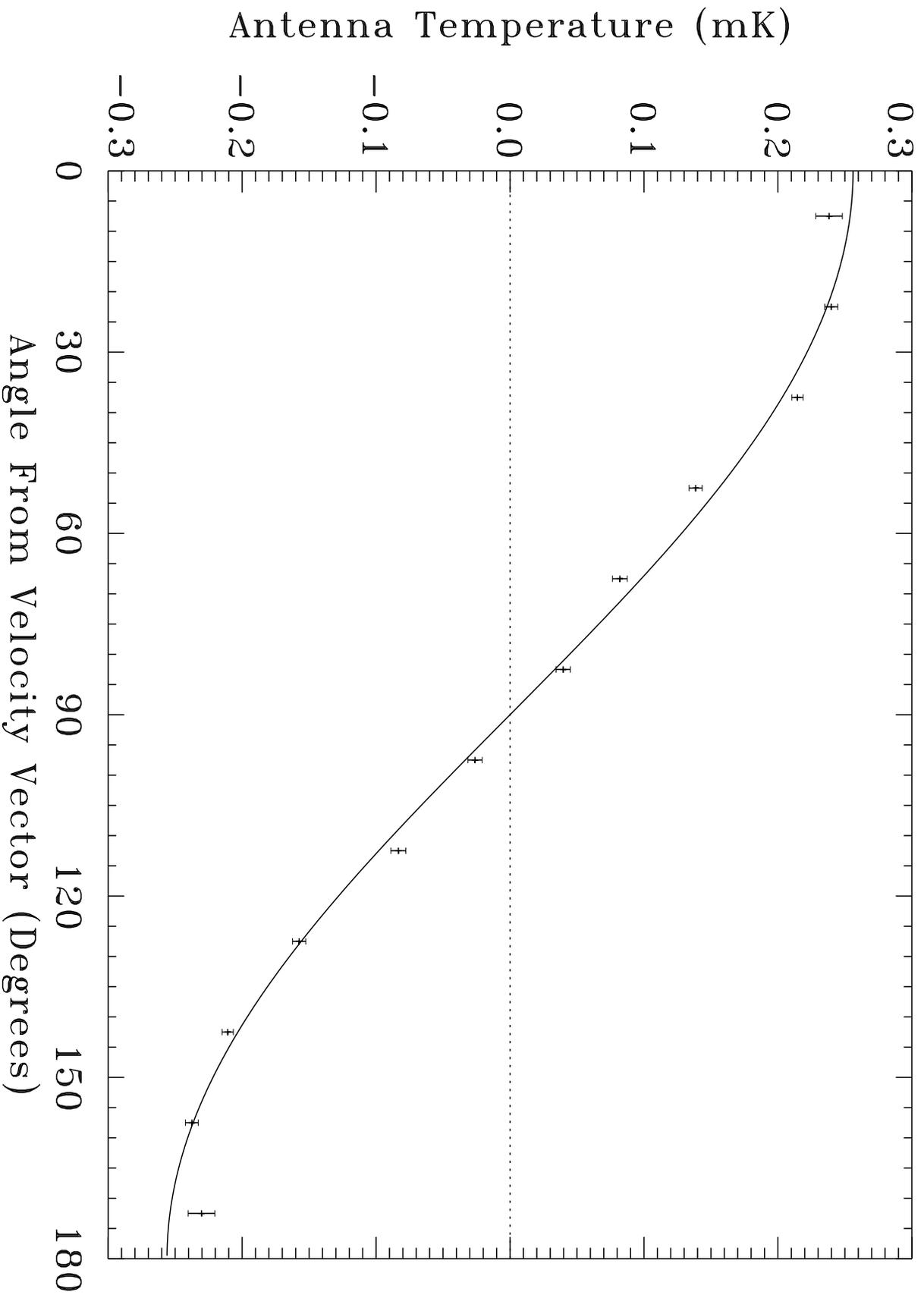

Figure 4